\documentclass[12pt]{article}
\usepackage{latexsym}
\usepackage{amssymb}
\usepackage{array}
\usepackage{graphicx}

\def\0{\over } \def\1{\vec } \def\2{{1\over2}} \def\4{{1\over4}}
\def\5{\bar } %\def\5{\overline }
\def\6{\partial }
\def\7#1{{#1}\llap{/}}
\def\8#1{{\textstyle{#1}}} \def\9#1{{\bf{#1}}}
\def\.{\cdot }
\def\^#1{\widehat{#1}}

\def\intx{\int\!d^3x\, }

\def\({\left(} \def\){\right)} \def\<{\langle } \def\>{\rangle }
\def\[{\left[} \def\]{\right]}  
 
\def\pmbf#1{\setbox0=\hbox{${#1}$}
        \kern-.025em\copy0\kern-\wd0
        \kern.05em\copy0\kern-\wd0
        \kern-.025em\raise.0433em\box0 }

\def\be{\begin{equation}}
\def\ee{\end{equation}}

\newcommand{\bel}[1]{\begin{equation}\label{#1}}
\def\bea{\begin{eqnarray}}
\newcommand{\beal}[1]{\begin{eqnarray}\label{#1}}
\def\eea{\end{eqnarray}}
\def\nn{\nonumber\\ }

\begin{document}

\begin{titlepage}
\renewcommand{\thefootnote}{\alph{footnote}}
~\vspace{-2.4cm}
\begin{flushright} 
TUW-05-02\\
YITP-SB-05-02\\ ITP-UH-04/05
\end{flushright}  
%\begin{center}  \vfil %\vspace{1.6cm}
%{%\large 
\bigskip
\vfil
\centerline{\Large\bf BPS saturation of the $N=4$ monopole
}
\medskip
\centerline{\Large\bf by infinite composite-operator renormalization}
\medskip
\begin{center}
\vfil 
{\large\sc
A. Rebhan$^1$\footnote{\footnotesize\tt rebhana@hep.itp.tuwien.ac.at}, 
R. Sch\"ofbeck$^1$\footnote{\footnotesize\tt schoefbeck@hep.itp.tuwien.ac.at},\\[3pt]
P. van Nieuwenhuizen$^2$\footnote{\footnotesize\tt vannieu@insti.physics.sunysb.edu} 
and
R. Wimmer$^3$\footnote{\footnotesize\tt wimmer@itp.uni-hannover.de}
}\\
\end{center}  \medskip \smallskip \qquad \qquad 
{\sl $^1$} \parbox[t]{12cm}{\sl 
  Institut f\"ur Theoretische Physik, Technische Universit\"at Wien, \\
  Wiedner Hauptstr. 8--10, A-1040 Vienna, Austria\\ } \\
\bigskip \qquad \qquad 
{\sl $^2$} \parbox[t]{12cm}{\sl 
  C.N.Yang Institute for Theoretical Physics, \\
  SUNY at Stony Brook, Stony Brook, NY 11794-3840, USA } 
\medskip \qquad \qquad 
{\sl $^3$} \parbox[t]{12cm}{\sl 
Institut f\"ur Theoretische Physik, Universit\"at Hannover,\\ 
Appelstr.~2, D-30167 Hanover, Germany\\ } \\
\vfil
\centerline{ABSTRACT}\vspace{.5cm}
Quantum corrections to
the magnetic central charge of the monopole in 
$N=4$ supersymmetric Yang-Mills
theory are free from the anomalous contributions that are crucial
for BPS saturation of 
the two-dimensional supersymmetric kink
and the $N=2$ monopole. However these
quantum corrections
are nontrivial and they
require infinite renormalization
of the supersymmetry current, central charges, and energy-momentum tensor,
in contrast to $N=2$ and
even though %(or rather because)
the $N=4$ theory is finite. Their
composite-operator renormalization
leads to counterterms which form a multiplet
of improvement terms. 
Using on-shell renormalization conditions
the quantum
corrections to the mass and the central charge then vanish both,
thus verifying quantum BPS saturation. % indeed holds.
\vfil

%{\scriptsize \hfill Version of \today}
\end{titlepage}

\setcounter{footnote}{0}

Solitons in supersymmetric (SUSY) quantum field theories 
\cite{D'Adda:1978mu,Witten:1978mh} (and further
references in \cite{Goldhaber:2004kn})
continue
to produce surprises. A new anomalous contribution
to the central charge of the 1+1-dimensional $N=1$ kink was needed
for BPS saturation 
\cite{Rebhan:1997iv,Nastase:1998sy,Graham:1998qq,Shifman:1998zy};
using supersymmetry-preserving dimensional regularization,
suitably adapted to take solitons into account \cite{Rebhan:2002uk},
this anomaly was due to a kink-induced left-right asymmetry of the momenta
of the fermion in the extra dimension \cite{Rebhan:2002yw}.
For the 2+1-dimensional $N=2$ vortex no such anomalies appeared,
but a finite quantum correction to the central charge, induced
by the winding of the background fields, was essential for
BPS saturation \cite{Rebhan:2003bu}.\footnote{For a superfield
treatment of the SUSY kink and the vortex see Refs.\
\cite{Fujikawa:2003gi,Shizuya:2003vm,Shizuya:2004ii} 
and \cite{Shizuya:2005th}, respectively.}
Recently, we studied quantum corrections to
the monopole in 3+1-dimensional $N=2$ SUSY Yang-Mills theory. While the
infinite renormalization of the coupling constant cancelled
a logarithmic divergence in the mass \cite{Kaul:1984bp}
and the central charge \cite{Imbimbo:1985mt}, we obtained
unexpected finite quantum corrections to both the
mass and the central charge \cite{Rebhan:2004vn} which are completely analogous
to the anomalous contributions of the $N=1$ kink
of Ref.~\cite{Rebhan:2002yw}, and in fact essential
for consistency with the $N=2$ low-energy effective action
of Seiberg and Witten 
\cite{Seiberg:1994rs,Seiberg:1994aj,Alvarez-Gaume:1997mv}.

In this Letter we consider the monopole in the $N=4$ SUSY
Yang-Mills model in 3+1 dimensions. Since this is a finite theory,
one expects no contributions from anomalies, nor infinite
counterterms from the renormalization
of physical parameters.%
%, at least in
%background covariant gauges where bosonic wave function renormalization
%shares the finiteness of the coupling constant renormalization.%
\footnote{However there are divergent
wave-function renormalizations in %non-background-covariant
Wess-Zumino
%non-Feynman 
gauge \cite{Kovacs:1999fx}, infrared-divergences in general supergauges \cite{Kovacs:1999fx},
and logarithmic divergences in Green functions of gauge-invariant
composite operators corresponding to unprotected long multiplets
\cite{Bianchi:1999ge}.
}
Thus the sum over zero point energies 
%(which is always evaluated in background covariant gauges)
should give a finite,
perhaps vanishing, one-loop quantum correction to the mass,
and this is indeed the case as shown in Ref.~\cite{Imbimbo:1985mt}.
Then also the central charge ought to be finite, and equal to
the mass to comply with BPS saturation \cite{Witten:1978mh}.
But comparison with the renormalization of infinities
in the $N=2$ case as worked out in Ref.~\cite{Imbimbo:1985mt}
produces a perplexing puzzle.%
\footnote{We thank S.\ Mukhi for drawing our attention to it.}
In the $N=2$ model the quantum
corrections to the central
charge in the monopole background
have an ultraviolet divergence which is cancelled
by the infinite renormalization of the coupling constant.
However, in the $N=4$ model the extra four scalar fields and
the additional fermions
do not contribute to the central charge, while
the coupling constant no longer requires a counterterm.
Hence, the one-loop central charge seems divergent,
while the one-loop contribution to the mass is zero, so that
the $N=4$ model in the sector with solitons,
in contrast to $N=2$, 
is not finite, %and appears %to violate the BPS bound.
%to be 
in apparent conflict with BPS saturation.

Confronted with this situation, one may consider various
possibilities for a solution: 

(i) that fermionic terms
in the central charge could contribute. Fermionic terms of course contribute
to the mass, but common lore has it that they do not contribute
to central charges and other integrals of total derivatives.
However, while the supersymmetry algebra involves numerous
fermionic terms (which are in fact responsible
for the anomalous contribution to the central charge
of the $N=2$ theory \cite{Rebhan:2004vn}), they disappear from central charges
in strictly four dimensions and thus should not be able
to give rise to infinities;

(ii) that one should add improvement terms to the currents.
%they yield a
%total derivative to the SUSY current. 
Since in the spontaneously broken phase one has a mass scale
(the vev of the Higgs scalar), it is doubtful that improved
currents should play a role.
Indeed, a SUSY variation of the improvement term to the SUSY
current
produces an extra bosonic total derivative in the central
charge current and in the energy density that would
modify mass and central charge already at the classical level.
If the SUSY algebra is to represent the monopole with
the standard value of mass and central charge,
the relevant SUSY current has to be unimproved, in the
form that is determined by dimensional reduction of the
10-dimensional theory;

(iii) that one should simply renormalize the central charge. This would solve
the problem of the finiteness of the central charge, but supersymmetry
would then  require that one also renormalize 
the SUSY current and the stress tensor. This would
seem to clash with the calculated finiteness of the quantum mass of the
monopole.

In this Letter we present a solution which is as surprising as
it is simple: if one considers the multiplet associated with
the SUSY current 
\cite{Ferrara:1974pz,Ogievetsky:1976qc,Sohnius:1978pk,Howe:1981qj,Bergshoeff:1980is,Bergshoeff:1982av}, the multiplet
of unimproved currents of the
$N=4$ model requires additive infinite renormalization involving
a multiplet of improvement currents as counter\-terms. 
%The underlying reason is that the multiplet of improved
%currents is superconformal and does not renormalize in $N=4$.
The $N=2$ model
differs by the absence of a 5-form 
%has fewer independent
charge in the higher-dimensional SUSY algebra,
which 
permits cancellations not possible in $N=4$,
and both the unimproved and the improved central charge density 
turn out to be finite
by themselves.

To demonstrate these features explicitly, we describe
the $N=4$ model in more detail and set up our notations.
The $N=4$ super Yang-Mills theory is most easily obtained
from trivial dimensional reduction from 9+1 dimensions
\cite{Brink:1977bc,Gliozzi:1976qd}.
We shall write $x^M$ for the coordinates and
$A^{a}_M$ for the ten-dimensional Yang-Mills field
with $M$ running over $0,\ldots 3,5,\ldots 10$
and $a$ the index for colour (for simplicity we shall
take SU(2)).
%The monopole will be located in the four Euclidean components
%$A_1,A_2,A_3,A_6$. -- A_5 \not=P of N=2 !
The action reads
\be
\mathcal L=-\4(F_{MN}^a)^2-{1\02}\bar\lambda^a \Gamma^M (D_M \lambda)^a
\ee
with a Majorana-Weyl spinor $\lambda$
and is invariant under $\delta A_M=-\bar\epsilon \Gamma_M \lambda$
and $\delta \lambda = \2 F_{MN} \Gamma^{MN} \epsilon$.

The SUSY current is $j^M=\2 F_{NP} \Gamma^{NP} \Gamma^M \lambda$
and varies under SUSY into
\bea\label{dj}
\2 \delta j^M(x) &=&
(F^{MP} F_{NP} - \4 \delta^M_N F^{RS} F_{RS} + \2 \bar\lambda
\Gamma^M D_N \lambda) \Gamma^N \epsilon \nn
&&+{1\016}(\bar\lambda \Gamma^M \Gamma^{PQ} D^R \lambda) \Gamma_{PQR}\epsilon\nn
&&+{1\08}F_{NP}F_{QR} \Gamma^{NPMQR} \epsilon
\eea
The term with $\Gamma_{PQR}$ vanishes after integration over $x$
since $\{Q,Q\}$ is symmetric in the two spinor indices.

For the purpose of dimensional reduction, the 16-component Majorana-Weyl
spinor $\lambda^a$ is written as $(\lambda^{\alpha Ia},0)$, where
$\alpha=1,\ldots,4$ is
the 4-dimensional spinor index, 
$I=1,\ldots,4$ is the rigid SU(4) index, and $a$ is the adjoint
SU(2) colour index.
For the gamma matrices we use the representation\footnote{We use
$\{\gamma_\mu,\gamma_\nu\}=2\eta_{\mu\nu}$ with metric signature
$-+++$, and $\mu,\nu=0,1,2,3$. Further, 
%$\gamma_5=\gamma^1\gamma^2\gamma^3i\gamma^0$,
$\gamma_5=i\gamma_0\gamma_1\gamma_2\gamma_3$,
so $\{\gamma_5,\gamma_\mu\}=0$
and $\gamma_5^2=\mathbf 1$.
Majorana spinors satisfy $\lambda^TC_{11}=\lambda^\dagger
i\Gamma^0$  in 9+1 dimensions
and $(\lambda^I)^TC_4=(\lambda^I)^\dagger i \gamma^0$
in 3+1 dimensions.}
\bea\label{Gammas}
\Gamma_\mu &=& \gamma_\mu \otimes {\bf 1} \otimes \sigma_2, 
\quad \mu=0,1,2,3,\nn
\Gamma_{4+j}&=& {\bf 1} \otimes \alpha_j \otimes \sigma_1, \nn
\Gamma_{7+j}&=& \gamma_5 \otimes \beta_j \otimes \sigma_2,
\quad j=1,2,3, \nn
\Gamma_{11}&=&{\bf 1} \otimes {\bf 1} \otimes \sigma_3,
\qquad C_{11}=-iC_4\otimes {\bf 1} \otimes \sigma_1.
\eea
The $\alpha_j$ and $\beta_j$ are the six generators of SO(3)$\times$SO(3)
in the representation of purely imaginary antisymmetric $4\times4$ %'t Hooft
matrices \cite{'tHooft:1976fv,Figueroa:EDC},
self-dual and anti-self-dual, respectively, and
satisfying $\{\alpha_i,\alpha_j\}=\{\beta_i,\beta_j\}=2\delta_{ij}$,
$[\alpha_i,\alpha_j]=2i\epsilon_{ijk}\alpha_k$,
$[\beta_i,\beta_j]=2i\epsilon_{ijk}\beta_k$,
and $[\alpha_j,\beta_k]=0$.

After reduction to 3+1 dimensions, the energy operator $T_{00}$
contains the sum of the stress tensors for the gauge fields $A_\mu^a$,
the four Majorana spinors $\lambda^{\alpha Ia}$, the
three scalars $S_j=A_{4+j}$ and the three pseudoscalars
$P_j=A_{7+j}$.

The 12 real central charges appear as\footnote{The 
complex matrix $Z^{IJ}=-Z^{JI}$ of central charges %with I,J=1,4
contains 6 complex (12 real) elements, as shown in (\ref{QQ}).
The magnetic charge $U_1$ and the electric charge $\tilde V_1$
only appear in the combination $Z^{IJ}=(iU_1+\tilde V_1)(\alpha^1)^{IJ}
+\ldots$.
By a unitary
transformation one can block-diagonalize $Z^{IJ}$, with two real antisymmetric
2$\times$2 matrices along the diagonal.
%The corresponding charges are the magnetic
%and electric charge, for example $U_1$ and $\tilde V_1$.
The $N=4$ action has a
rigid $R$ symmetry group SU(4) (not U(4) \cite{Sohnius:1985qm}).
To exhibit this SU(4), the spin zero fields
are combined into $M^{IJ}=(\alpha^j)^{IJ}S_j +i (\beta^j)^{IJ}P_j$, 
but if $S_1^a=v\delta^a_3$, there is a
central charge and $R$ is broken down to the manifest stability subgroup of
$M^{23}=M^{14}=S_1$,
which is USp(4).}
\bea\label{QQ}
\2\{ Q^{\alpha I},Q^{\beta J} \} &=&
\delta^{IJ} (\gamma^\mu C^{-1})^{\alpha\beta} P_\mu \nn
&+&\!\!\!i(\gamma_5 C^{-1})^{\alpha\beta} (\alpha^j)^{IJ} \intx U_j
-(C^{-1})^{\alpha\beta} (\beta^j)^{IJ} \intx V_j \nn
&+&\!\!\!(C^{-1})^{\alpha\beta} (\alpha^j)^{IJ} \intx \tilde V_j
+i(\gamma_5 C^{-1})^{\alpha\beta} (\beta^j)^{IJ} \intx \tilde U_j\,,
\eea
where $U_j$ and $V_j$ are due to the five-gamma term in (\ref{dj}),
and $\tilde U_j$ and $\tilde V_j$ due to the one-gamma terms.
The indices $I$ and $J$ are lowered and raised by the charge conjugation
matrix in this space, which is $\delta^{IJ}$, see (\ref{Gammas}).

The $N=2$ monopole (and dyon) can be embedded 
into the $N=4$ model by selecting $j=1$
and one then has ($S=S_1$, $P=P_1$, $U=U_1$ etc.)% %\cite{Imbimbo:1985mt}
\footnote{To obtain the total derivatives in (\ref{UUVV}),
one needs to use Bianchi identities in the case of $U$ and $V$, and
equations of motion in the case of $\tilde U$ and $\tilde V$.}
\bea\label{UUVV}
&&U=\6_i(S^a \2 \epsilon^{ijk} F_{jk}^a),
\qquad \tilde U=\6_i(P^a F_{0i}^a) \nn
&&V=\6_i(P^a \2 \epsilon^{ijk} F_{jk}^a),
\qquad \tilde V=\6_i(S^a F_{0i}^a).
\eea

The classical contribution to the central charge of a magnetic monopole
is due to $U$, and it is determined by the asymptotic values
\bea
A_i^a \to \epsilon_{aij}{1\0g}{\hat x^j\0r},\qquad
F_{ij}^a\to - \epsilon_{ijk}{1\0g}{\hat x^a \hat x^k\0r^2},\nn
S^a_1\to - \delta_i^a \hat x^i (v-{1\0gr}),\qquad \hat x^i \equiv x^i/r,
\eea
yielding $\intx U=4\pi v/g$.

For the above realization of central charges it is crucial
that the SUSY current $j$ is exactly as determined by the
9+1-dimensional theory and does not have improvement terms
at the classical level.
In 3+1 dimensions one could add a gauge invariant improvement
term of the form
\be\label{Dj}
\Delta j^\mu=c\Gamma^{\mu\nu}\6_\nu (A_{\mathcal J} \Gamma^{\mathcal J} \lambda),
\quad {\mathcal J}=5,\ldots,10,
\ee
which for $c=-2/3$ would provide $\gamma_\mu (j+\Delta j)^\mu=0$.
However, this would change $U$ into $U+\Delta U$ with
\be\label{DU}
\Delta U= {c\02} \left(U+{i\08}\6_i (\epsilon^{ijk}
\bar \lambda \alpha^1 \gamma_{jk} \lambda) \right)
\ee
and would change the value of the central charge $\intx U$ at the classical
level. Similarly, the stress tensor would be modified, resulting
in a corresponding shift of mass ascribed to the monopole.

%We now discuss some renormalization properties...
For the calculation of quantum corrections we need to fix the gauge.
The most convenient choice turns out to be the background-covariant
Feynman-$R_\xi$ gauge (with $\xi=1$)
obtained by dimensional reduction of the gauge
fixing term 
\be\label{Lgf}
\mathcal L_{g.f.}=-{1\02\xi}(D_M[\hat A]\,a^M)^2
\ee
where $a^M$ are the quantum gauge
fields and $\hat A^M$ the background fields.

To determine potential counter\-terms, it is sufficient to consider
the spontaneously broken phase in the trivial background
with no monopoles. Choosing
$\langle S_1^a \rangle = v \delta^a_3$, one finds that
tadpole diagrams with an external background field $\hat S_1$ 
or an external quantum field $s_1$ vanish.
Coupling constant renormalization can be fixed by wave-function
renormalization of the gauge boson background fields due to $Z_g=Z_A^{-1/2}$.
Using the on-shell two-point self-energy of the massless photon
%or the massless Higgs boson
with external background fields, one finds after a Wick rotation
\be\label{g2ren}
{1\0g_0^2}={1\0g^2}+c_g I,\quad
I\equiv\int {d^{4+\epsilon}k_E\0(2\pi)^{4+\epsilon}}
{1\0 (k^2_E + m^2)^2 },\quad m=g_0 v_0=gv,
\ee
with $c_g=4$ in $N=2$, but $c_g=0$ in the present case of $N=4$.
Renormalizing similarly the massless Higgs boson on-shell,
one finds $Z_S=Z_A$ in the gauge $\xi=1$, %(and only then).
but 
\be\label{ZS}
Z_S=(1+2g^2[4-N+1-\xi]I)\ee
for $\xi\not=1$, while $Z_A$ is $\xi$-independent.

The one-loop result for the mass of a monopole is given by
\be
M^{(1)}={4\pi m\0g_0^2}+\2\sum(\omega_B-\omega_F).
\ee
In the $N=2$ case the sum over zero-point energies 
gives rise to an integral over a nontrivial difference of spectral
densities which is UV divergent such that the infinities cancel
with those of (\ref{g2ren}), leaving a finite remainder,
namely the anomalous contribution discussed in Ref.~\cite{Rebhan:2004vn}.
In $N=4$, however, there is complete cancellation of the
sums over zero-point energies \cite{Imbimbo:1985mt}.
The requirement of quantum BPS saturation thus demands a
nontrivial contribution to the central charge in the case of $N=2$,
but none for $N=4$.

In Ref.~\cite{Rebhan:2004vn},
the anomalous contribution to the central charge for $N=2$ was obtained
as a fermionic contribution to the momentum operator in the
extra dimension required for supersymmetry-preserving dimensional regularization.
This involved a difference of spectral densities for the different
components of the fermionic fluctuations. In the case of $N=4$,
there are twice as many fermions, and it turns out that 
the extra fermions lead to a complete cancellation of anomalous contributions.

The non-anomalous quantum corrections to the central charge 
of %the $N=4$ 
a mono\-pole can be
calculated along the lines of Ref.~\cite{Imbimbo:1985mt}. At one-loop order
they are given by
\be
U^{(1)}_{non-ano}={4\pi m\0g_0^2}+
\2\int d^3 x\, \6_i \left(
\hat S^a \epsilon_{ijk} \langle
F^a_{jk}[\hat A+a]-F^a_{jk}[\hat A] \rangle \right).
\ee
Expanding in terms of $a$,
the term quadratic in $a$ produces a loop formed by a propagator for
the $a_\mu^a$ field in the monopole background, whose spatial components
in background-covariant Feynman-$R_\xi$ gauge read \cite{Imbimbo:1985mt}
\be
\langle a_j^b(x) a_k^c(y) \rangle
= i [(\bar D^2)^{-1bc}\delta_{jk}
+2(\bar D^2)^{-1bb'}g\epsilon^{b'c'e}\hat F_{jk}^e
(\bar D^2)^{-1c'c}
+\ldots] \delta^{4+\epsilon}(x-y)
\ee
where $(\bar D^2)^{-1}$ is the dimensionally reduced version of 
$(D_M[\hat A] D^M[\hat A])^{-1}$ with
asymptotic behaviour
\be
(\bar D^2)^{-1ab} \to (\Box - m^2+i\epsilon)^{-1}(\delta^{ab}
-\hat x^a \hat x^b)+ (\Box+i\epsilon)^{-1} \hat x^a \hat x^b.
\ee
Using that $\hat S^a\to - v \hat x^a$ and that
to leading order in $1/r$ the vector $\hat x^a$ can be treated as constant,
one readily finds \cite{Imbimbo:1985mt}
\be\label{U1na}
U^{(1)}_{non-ano}={4\pi m\0g_0^2}+16\pi i m \lim_{x\to y}
{1\0(\Box-m^2+i\epsilon)^2}\delta^{4+\epsilon}(x-y).
\ee
In $N=2$, where $c_g=4$ in (\ref{g2ren}), this leads to $U^{(1)}_{non-ano}
=U^{(1)}_{cl}=4\pi m/g^2$.
However, in $N=4$ the coupling constant does not renormalize, $g_0=g$,
and we are left with an uncancelled ultraviolet divergence in $U^{(1)}$.

As we already discussed, the solution to this problem does not come from
possible
fermionic contributions to the central charge 
density $U$ in $N=4$, as there are none.
Using the improved SUSY current and
the corresponding central charge would in fact change the situation,
but then mass and central charge would not agree with the
standard value at the classical level.
%It also does not help to consider instead the improved SUSY current and
%the corresponding central charge, since this would not give
%the correct value already at the
%classical level. 
Keeping the unimproved SUSY current as obtained by dimensional reduction,
the uncancelled divergence in (\ref{U1na}) has instead
to be absorbed in a composite-operator renormalization of $U$.

\begin{figure}[t]
\centerline{\includegraphics[angle=-90,scale=0.59,bb=110 110 420 775]{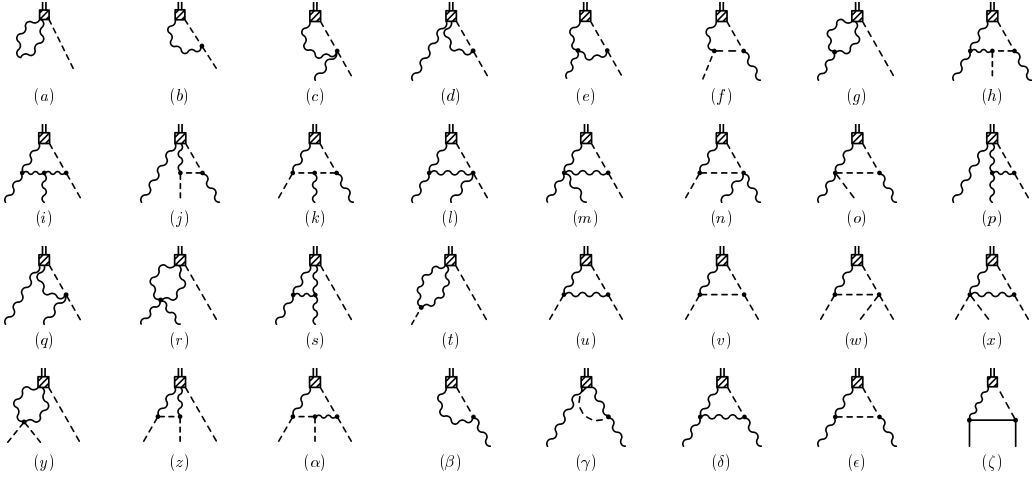}}
\caption{One-particle-irreducible divergent
one-loop diagrams in matrix elements
involving the central charge density $U$. 
Wavy lines indicate gauge bosons, dashed lines scalar bosons,
and full lines fermions.\label{figU}}
\end{figure}

This perhaps surprising result can already be seen in the renormalization
of the $N=4$ theory in the trivial sector (i.e.\ without monopoles).
To discuss composite-operator renormalization we consider
the partition function with an external source $K$ for the composite operator
$U$ defined in (\ref{UUVV}), $\mathcal L \to \mathcal L+K(x) U(x)$.
All additional ultraviolet divergences that appear in matrix
elements with elementary fields (or further insertions of $U$)
have to be absorbed by counter\-terms linear in $K$ (or of higher
order in $K$)\footnote{In much present work on the AdS-CFT correspondence,
the divergences in correlation functions with two or more
composite (gauge invariant) operators at nearby points were
considered \cite{D'Hoker:1998tz,Eden:1998hh,Chalmers:1999gc,Arutyunov:2000ku,D'Hoker:1999jp}.}.
For evaluating $\langle U \rangle$, which will of course
give something nontrivial only in the topologically nontrivial sector,
we only need counter\-terms linear in $K$.
These are determined by the 32 proper diagrams in Fig.~\ref{figU}, which list
the potential divergent matrix elements with elementary fields.
%(disregarding 1-particle-reducible diagrams).
Of particular interest to us are those involving external scalars $S_1^a$
and gauge fields. The superficially quadratically divergent diagrams
(\textit a) and (\textit b) vanish identically. 
The diagrams (\textit c)--(\textit g) are superficially linearly divergent, 
but if the external scalar is the massless Higgs field
$S_1^3$ and the external gauge field the massless photon $A_\mu^3$,
diagram (\textit c) vanishes in background-covariant Feynman-$R_\xi$ gauge.
Diagrams (\textit d)--(\textit f) turn out to be logarithmically
divergent individually, but finite when added together.
Diagrams (\textit t)--($\epsilon$) vanish for external $S_1^3$ and
$A_\mu^3$ fields.
The only remaining divergent diagram with these external fields 
%with external $S_1^3$ and
%$A_\mu^3$ 
is diagram
(\textit g), which
%Taking the external scalar to be the massless Higgs field
%$S_1^3$ and the external gauge field the massless photon $A_\mu^3$,
%%a straightforward calculation %
%evaluation in background-covariant Feynman-$R_\xi$ gauge
gives the logarithmically divergent
contribution 
\be\label{diagg}
4ig^2 q^i \epsilon^{ijk}p^j\delta_\mu^k
\int {d^{4+\epsilon}k\0(2\pi)^4} {1\0(k^2+m^2-i\epsilon)[(k-p)^2+m^2-i\epsilon]}
\ee
where $q$ is the incoming momentum carried by $U$ and $p$ the 
outgoing momentum
of $A_\mu^3$.
Thus the required counter\-term is proportional to
\be
K(x)\6_i(\epsilon_{ijk}\6_j \hat A_k^3(x) \hat S_1^3(x)).
\ee
The diagrams (\textit h)--(\textit s) are also logarithmically divergent
and evidently needed for making the counter\-term gauge invariant, namely
proportional to $K(x)\hat U(x)$.

%In $N=2$ super-Yang-Mills theory, such a counter\-term with
%the required prefactor is already
%provided by wave-function and coupling renormalization
%of $\hat U(x)$, so it does not need to be added.
%However, in the case of $N=4$, where the above calculation does not
%involve the additional scalars and fermions of this theory,
In the $N=4$ theory with background-covariant Feynman-$R_\xi$ gauge,
there are no divergent counter\-terms from bosonic wave function or
coupling renormalization, so $U(x)$
needs to be renormalized explicitly. With the on-shell renormalization
condition that diagram (\textit g) is subtracted completely at $p=0$, 
the counter\-term is such that the second term
in (\ref{U1na}) is cancelled completely, as is required for
BPS saturation at the quantum level. 

%The diagrams (\textit t)--($\zeta$) 
%are also superficially logarithmically divergent
%and correspond to operator mixing (counter\-terms linear in $K$ but
%involving other operators than $U$). These operators are also
%total derivatives, but they do not contribute in
%the calculation of $\langle U \rangle$.

%We thus conclude that the monopole of the finite $N=4$ super Yang-Mills theory
%does not receive quantum corrections to its mass at the one-loop
%level, but the central charge operator does and in fact requires
%infinite composite-operator renormalization.
%This is in curious contrast to the $N=2$ monopole, where
%the central charge does not require renormalization beyond
%elementary renormalizations of coupling and wave functions.

However, the central charge density $U(x)$
is not just multiplicatively renormalized.
Consider diagram ($\zeta$); it yields
\be\label{diagz}
%-4i
4ig^2\!\int {d^{4+\epsilon}k\0(2\pi)^4} {q^i\epsilon^{ijk}
k^jk^s\alpha^1\gamma_{ks}
\0 (k^2+m^2-i\epsilon)((p+k)^2+m^2-i\epsilon)((k+p+q)^2+m^2-i\epsilon)}.
\ee
Renormalizing at vanishing external momenta, one can replace
$k^j k^s$ by $\eta^{js}k^2/4$.
% and the resulting
%fermionic counter\-term corresponds to the
%second term in (\ref{DU}). 
Taking into account that
the fermionic counterterm has to include a factor $1/2$ because
of the Majorana property, the complete counterterm required by
the composite-operator renormalization of the central charge 
density turns out to
be proportional to the improvement term $\Delta U$ and explicitly reads
\be\label{N4ct}
4g^2 I K(x)\left(\hat U+{i\08}\6_i (\epsilon^{ijk}
\bar \lambda \alpha^1 \gamma_{jk} \lambda) \right)\!(x)
\ee
where $I$ is the divergent integral appearing in (\ref{g2ren}).
This result is in fact independent of the gauge parameter $\xi$
in (\ref{Lgf}). While the gauge-parameter independence of diagram ($\zeta$)
is fairly easy to check, 
the purely bosonic diagrams (\textit c)--(\textit g) 
are gauge dependent
such that it compensates for the gauge dependence of $Z_S$
displayed in (\ref{ZS}).\footnote{Explicitly, for $\xi\not=1$
the factor 4 in front of (\ref{diagg}) gets replaced
by $5-\xi$ which arises as
$0+4+(-2\xi)+(-2)+(3+\xi)$
from the five diagrams (\textit c)--(\textit g), respectively.
The factor 4 in the final
counterterm (\ref{N4ct}) is recovered by 
%taking into account that
adding to minus the contributions from these
diagrams the contribution from
$Z_S^{1/2}=(1+g^2(1-\xi)I)$ in $N=4$.}

The additive renormalization of $U$ can be understood as follows.
Writing the unimproved SUSY current as $j=(j+\Delta j)-\Delta j
\equiv j_{improved}-\Delta j$, 
with $\Delta j$ given by (\ref{Dj}) and $c=-2/3$ therein,
it is the superconformal
current \cite{Bergshoeff:1980is}
$j_{improved}$ which is protected from renormalization
\cite{Hagiwara:1979pu}, but
the multiplet of improvement terms associated to $\Delta j$ is not
(compare with the renormalization of an unimproved
energy-momentum tensor \cite{Callan:1970ze,Freedman:1974gs,Freedman:1974ze,Collins:1976vm,Brown:1980pq}).
In the presence of monopoles, the relevant SUSY current is
the unimproved one, hence the appearance of infinite renormalizations.

To verify that the improved SUSY current and its superpartners
are not renormalized while the improvement terms are
multiplicatively renormalized, one in fact needs to take into account
the wave function renormalization of the fermions. 
Regardless of background covariance the latter is not
related to the wave function renormalization of the gauge bosons,
since the gauge fixing term (\ref{Lgf}) is not SUSY
invariant. A straightforward calculation of the one-loop
fermion self-energy, renormalized such that it vanishes
on-shell for the massless fermions, leads to
\be\label{Zpsi}
\lambda_0=\sqrt{Z_\lambda}\lambda,\qquad
Z_\lambda=1-2 (N+\xi-1) g^2 I
\ee
%for $\lambda_0=\sqrt{Z_\lambda}\lambda$
in $N$-extended super-Yang-Mills theory.
%so this is divergent even for $\xi=1$ and $N=4$.

\begin{figure}[t]
\centerline{\includegraphics[scale=0.65,bb=125 630 450 705]{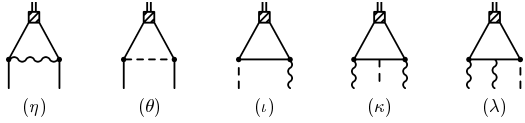}}
\caption{One-particle-irreducible divergent
one-loop diagrams in matrix elements
involving the fermionic term in the improvement term %$\Delta U$, eq.\
(\ref{DU}). 
\label{figDUf}}
\end{figure}

Considering now the renormalization of the fermionic composite
operator in $\Delta U$, eq.~(\ref{DU}), this is determined 
by the diagrams of Fig.~\ref{figDUf}.
Evaluating diagrams $(\eta)$ and $(\theta)$, which can
give rise to an extra fermionic counterterm, one finds
that these contribute proportional to $(\xi-1)$ such
that the fermionic operator in
$\Delta U$ is renormalized by a gauge-choice independent
counterterm when subtracted on-shell.

The remaining diagrams ($\iota$)--($\lambda$) are
trivially gauge-parameter independent and lead to
mixing with the bosonic operator $U$ precisely such that ${\Delta U}$
is multiplicatively renormalized as claimed. The counterterm
is given explicitly by
\be\label{ZDeltaU}
Z_{\Delta U}=1+12g^2I
\ee
in $N=4$.

In the $N=2$ theory, the situation is subtly different.
Because of the absence of a 
5-form charge in the higher-dimensional
SUSY algebra,
the central charge of the $N=2$ theory is given by the combination
\be\label{U2full}
U_{N=2}=\2 \epsilon^{ijk} \6_i(S^a F_{jk}^a)
+\6_i(P^aF_{i0}^a) = U-\tilde U.
\ee
Both $U$ and $\tilde U$ give rise to fermionic counterterms
through diagram ($\zeta$) of Fig.~\ref{figU}, but they turn out
to be equal and cancel when combined.

Correspondingly, one finds that the improvement 
term $\Delta j^\mu$, which is given by
(\ref{Dj}) with $\mathcal J$ running only over $\mathcal J=5,6$, now 
does not give rise to a fermionic term
in $\Delta U_{N=2}$. Instead, the latter has the same structure
as $U_{N=2}$ itself:
\be\label{DU2}
\Delta U_{N=2}={c\02} (U-\tilde U).
\ee

Moreover, in the $N=2$ case, the bosonic counterterms needed to
make $U_{N=2}$ finite are already contributed by the elementary
wave-function renormalizations of the bosonic fields,
and this equally holds for $\Delta U_{N=2}$. Hence,
$Z_{\Delta U}=1$ in $N=2$.

To summarize, while in the case of $N=2$ super-Yang-Mills theory
both the improved and the unimproved central charge density are
finite operators,
we have shown that 
in the $N=4$ case
composite-operator renormalization
of the central charge density is essential for obtaining
a finite result 
%in the background-covariant Feynman-$R_\xi$ gauge.
%in a calculation along the lines of Ref.~\cite{Imbimbo:1985mt}.
for the central charge at the quantum level.
Using on-shell renormalization conditions this then
matches with the null result obtained from calculating the
quantum mass of the $N=4$ monopole through the sum over
zero-point energies. It would be interesting to also calculate
the mass of the monopole from the expectation value of $T_{00}$,
which by supersymmetry receives also infinite additive
renormalization by an improvement term. The latter
is given by the total divergence $2
g^2  I \6_i\6_i(A_{\mathcal J}A^{\mathcal J})$,
which in fact does contribute to the mass of the soliton, since
in the monopole background
$A_{\mathcal J}A^{\mathcal J}=S^2\sim v^2(1-2/(mr))$ 
asymptotically. Because the sum over zero-point energies
can be obtained from the integral over $T_{00}$ by partial integrations,
which also give rise to total derivatives 
of the form of the improvement terms, we
expect that the composite-operator renormalization of $T_{00}$
is required to compensate the effect of the former.

We conclude that the resolution of the UV-divergence puzzle 
in the BPS saturation of the $N=4$ monopole has brought to the surface
a new feature, which did not play a role in the %lower-dimensional
models studied before. The multiplet of $N=4$ currents
for the theory with monopoles contains unimproved currents
which are not multiplicatively renormalized, but one must
add divergent counter\-terms which form a multiplet of $N=4$
improvement terms. 
We have found that these counter\-terms are not only
gauge invariant, but also independent of the
choice of gauge fixing parameter\footnote{%
It is still conceivable that a supersymmetric gauge choice could
eliminate these divergences.
%as in the unbroken theory
%\cite{Kovacs:1999fx}, 
However, explicit calculations of quantum corrections in the
presence of a monopole have been possible so far only
in ordinary background-covariant Feynman-$R_\xi$ gauge, and no
superfield version of that is known \cite{Goldhaber:2004kn}.},
which could imply that they have some wider importance.

\subsection*{Acknowledgments}
We would like to thank S.\ Mukhi
for very useful correspondence, and L.\ Rastelli, M. Ro\v cek, and 
M.\ Spradlin for discussions.
This work has been supported in part by the Austrian Science Foundation
FWF, project no. P15449.
P.v.N.\ and R.W.\ gratefully
acknowledge financial support from the International
Schr\"od\-inger Institute for Mathematical Physics, Vienna, Austria.
R.W.\ especially thanks Ines Schmiedmaier and Mathias Sadjadi for their
warm hospitality during his stay in Vienna.

\subsection*{Note added}

In the meantime we have performed a direct calculation
of the quantum mass of the $N=2$ and $N=4$ monopole from the
integral over the expectation value of $T_{00}$. We confirmed that the
total derivative terms, which result if one rewrites
the integral as a sum over zero point energies, indeed
give rise to ultraviolet divergent contributions in
the case of the $N=4$ monopole (but not in $N=2$)
which are cancelled by an improvement counterterm
with coefficient determined by (\ref{ZDeltaU}). The details of this calculation
will be presented elsewhere \cite{forthc}.

We note again that
our results depend on having chosen the unimproved
SUSY current multiplet that is singled out by the
possibility of formulating the $N=2$ and $N=4$ models
as dimensional reductions of $N=1$ super Yang-Mills theory
in $D=6$ and 10, respectively. In the $N=4$ model
the monopole mass would be different already at the classical level, but 
free from the necessity of composite operator renormalization
of the current multiplet, if one started from an improved
current multiplet. We believe that both formulations
are theoretically consistent, hence only experiment
could decide which one Nature chooses (if any).

\small
%\bibliographystyle{elsart-numwot}
%\bibliography{qft,ar,books}

\end{document}